\documentclass[manuscript]{acmart}

\copyrightyear{2023}
\acmYear{2023}
\setcopyright{rightsretained}
\acmConference[RecSys '23]{Seventeenth ACM Conference on Recommender Systems}{September 18--22, 2023}{Singapore, Singapore}
\acmBooktitle{Seventeenth ACM Conference on Recommender Systems (RecSys '23), September 18--22, 2023, Singapore, Singapore}\acmDOI{10.1145/3604915.3608845}
\acmISBN{979-8-4007-0241-9/23/09}



\usepackage{booktabs} 
\usepackage{color}
\usepackage{arydshln} %

\newcommand{\centeredcell}[1]{\begin{tabular}{l} #1 \end{tabular}}

\definecolor{cadmiumgreen}{rgb}{0.0, 0.42, 0.24}

\acmSubmissionID{582}

\begin{document}

\title[LLMs are Competitive Near Cold-start Recommenders]{Large Language Models are Competitive Near Cold-start Recommenders for Language- and Item-based Preferences}

\author{Scott Sanner}
\email{ssanner@mie.utoronto.ca}
\affiliation{%
  \institution{University of Toronto}
  \city{Toronto}
  \country{Canada}
}
\authornote{Work done while on sabbatical at Google.}
\author{Krisztian Balog}
\email{krisztianb@google.com}
\affiliation{%
  \institution{Google}
  \city{Stavanger}
  \country{Norway}
}
\author{Filip Radlinski}
\email{filiprad@google.com}
\affiliation{%
  \institution{Google}
  \city{London}
  \country{United Kingdom}
}
\author{Ben Wedin}
\email{wedin@google.com}
\affiliation{%
  \institution{Google}
  \city{Cambridge, MA}
  \country{United States}
}
\author{Lucas Dixon}
\email{ldixon@google.com}
\affiliation{%
  \institution{Google}
  \city{Paris}
  \country{France}
}

\begin{abstract}
Traditional recommender systems leverage users' item preference history to recommend novel content that users may like.  However, modern 
dialog interfaces that allow users to express language-based preferences offer a fundamentally different modality for preference input.  Inspired by recent successes of prompting paradigms for large language models (LLMs), we study their use for making recommendations from both item-based and language-based preferences in comparison to state-of-the-art item-based collaborative filtering (CF) methods.
To support this investigation, we collect a new dataset consisting of both item-based and language-based preferences elicited from users along with their ratings on a variety of (biased) recommended items and (unbiased) random items.  
Among numerous experimental results, we find that %
LLMs provide competitive recommendation performance for \emph{pure language-based preferences} (no item preferences) in the near cold-start case 
in comparison to item-based CF methods,
despite having no supervised training for this specific task (zero-shot) or only a few labels (few-shot). %
This is particularly promising as language-based preference representations are more explainable and scrutable than item-based or vector-based representations.

\end{abstract}

\maketitle

\section{Introduction}

\label{sec:intro}

The use of language in recommendation scenarios is not a novel concept. Content-based recommenders have been utilizing text associated with items, such as item descriptions and reviews, for about three decades~\citep{lops2011recsysbook}.
However, recent advances in conversational recommender systems have placed language at the forefront, as a natural and intuitive means for users to express their preferences and provide feedback on the recommendations they receive~\citep{gao-convrec-survey,jannach-convrec-survey}.
Most recently, the concept of natural language (NL) user profiles, where users express their preferences as NL statements 
has been proposed~\citep{Radlinski:2022:SIGIR}.  The idea of using text-based user representations is appealing for several reasons: it provides full transparency and allows users to control the system’s personalization. Further, in a (near) cold-start setting, where little to no usage data is available, providing a NL summary of preferences may enable a personalized and satisfying experience for users.
Yet, controlled quantitative comparisons of such NL preference descriptions against traditional item-based approaches are very limited.
Thus, the main research question driving this study is the following: How effective are prompting strategies with large language models (LLMs) for recommendation from natural language-based preference descriptions in comparison to collaborative filtering methods based solely on item ratings?

We address the task of \emph{language-based item recommendation} by building on recent advances in LLMs and prompting-based paradigms that have led to state-of-the-art results in a variety of natural language tasks, and which permit us to exploit rich positive and negative descriptive content and item preferences in a unified framework.  We contrast these novel techniques with traditional language-based approaches using information retrieval techniques~\cite{Balog:2021:SIGIR} as well as collaborative filtering-based approaches~\cite{Gantner:2011:MyMediaLite,ease}.
Being a novel task, there is no dataset for language-based item recommendation. As one of our main contributions, we present a data collection protocol and build a test collection that comprises natural language descriptions of preferences as well as item ratings. %
In doing so, we seek to answer the following research questions:
\begin{itemize}
    \item {\bf RQ1:} Are preferences expressed in natural language sufficient as a replacement for items for (especially) near cold-start recommendation, and how much does performance improve when language is combined with items? 
    \item {\bf RQ2:} How do LLM-based recommendation methods compare with item-based collaborative filtering methods? 
    \item {\bf RQ3:} Which LLM prompting style, be it completion, instructions, or few-shot prompts, performs best?
     \item {\bf RQ4:} Does the inclusion of natural language \emph{dis}preferences improve language-based recommendation?
\end{itemize}

\noindent Our main contributions are %
(1) We devise an experimental design that allows language-based item recommendation to be directly compared with state-of-the-art item-based recommendation approaches, and present a novel data collection protocol (Section~\ref{sec:expsetup});
(2) We propose various prompting methods for LLMs 
    for the task of language-based item recommendation (Section~\ref{sec:methods});
(3) We experimentally compare the proposed prompt-based methods against a set of strong baselines, including both text-based and item-based approaches (Section~\ref{sec:results}).
Ultimately, we observe that LLM-based recommmendation from pure language-based preference descriptions provides 
a competitive near cold-start recommender system that is based on an explainable and scrutable language-based preference representation.
\section{Related Work}

\emph{\textbf{Item-Based Recommendation.}}
Traditional recommender systems rely on item ratings. For a new user, these can be provided over time as the user interacts with the recommender, although this means initial performance is poor. Thus, preferences are often solicited with a questionnaire for new users \cite{Hu:2013:Interview, Rokach:2012:Initial,Sepliarskaia:2018:Optimized}. There has also been work looking at other forms of item-based preferences such as relative preferences between items \cite{Rokach:2012:Initial,Christakopoulou:2016:Towards}, although approaches that rely on individual item ratings dominate the literature.

Given a corpus of user-item ratings, very many recommendation algorithms exist. These range from methods such as item-based k-Nearest Neighbors \cite{Sarwar:2001:ICF}, where simple similarity to existing users is exploited, to matrix factorization approaches that learn a vector representation for the user \cite{Hu:2008:CFI,Ning:2011:SSL}, through to deep learning and autoencoder approaches that jointly learn user and item vector embeddings \cite{He:2017:Neural,Liang:2018:Variational,Huiyuan:2022:Denoising}. Interestingly, the EASE algorithm \cite{ease} is an autoencoder approach that has been found to perform on par with much more complex state-of-the-art approaches.

\emph{\textbf{Natural Language in Recommendation.}}
Following the proposals in \cite{Balog:2019:SIGIR,Radlinski:2022:SIGIR} to model preferences solely in scrutable natural language, recent work has explored the use of tags as surrogates for NL descriptions with promising results~\cite{mysore:2023:editable}.
This contrasts with, for instance \citet{hou2022towards}, who input a (sequence) of 
natural language item descriptions into an LLM to produce an (inscrutable) user representation for recommendation.
Other recent work has sought to use rich, descriptive natural language as the basis for recommendations.  At one extreme, we have narrative-driven recommendations~\citep{Bogers:2017:narrative-driven} that assume very verbose descriptions of specific contextual needs.  In a similar vein, user-studies of NL use in recommendation~\citep{Kang:2017:RecSys} identify a rich taxonomy of recommendation intents and also note that speech-based elicitation is generally more verbose and descriptive than text-based elicitation.
In this work, however, we return to the proposal in \cite{Radlinski:2022:SIGIR} and assume the user provides a more general-purpose language-based description of their preferences and dispreferences for the purpose of recommendation.

Recently, researchers have begun exploring use of language models (LMs) for recommendation tasks \cite{friedman2023leveraging}. %
\citet{Radlinski:2022:SIGIR} present a theoretical motivation for why LLMs may be useful for recommendations and provide an example prompt, but do not conduct any quantitative evaluation.
\citet{mysore2023large} generate preference narratives from ratings and reviews, using the narratives to recommend from held-out items.  \citet{penha2020does} show that off-the-shelf pretrained BERT~\cite{BERT} contains both collaborative- and content-based knowledge about items to recommend.  
They also demonstrate that BERT outperforms information retrieval (IR) baselines for recommendation from language-based descriptions.  However, they do not assess the relative performance of language- vs. item-based recommendation from LMs (for which we curate a dataset specifically for this purpose), nor does BERT's encoder-only LM easily permit doing this in a unified prompting framework that we explore here.
RecoBERT~\cite{malkiel2020recobert} leverages a custom-trained LM for deriving the similarity between text-based item and description pairs, with the authors finding that this outperforms traditional IR methods.   
\citet{hou2023large} focus on item-based recommendation, with an in-context learning (ICL) approach similar in spirit to our item-only few-shot approach. 
Similarly, \citet{kang2023llms} use an LLM to predict ratings of target items.
Finally, ReXPlug~\cite{hada2021rexplug} exploits pretrained LMs to produce explainable recommendations by generating synthetic reviews on behalf of the user.  None of these works, however, explore \emph{prompting strategies} in large LMs to \emph{translate actual natural language preferences into new recommendations} compared directly to item-based approaches. 

Further, we are unaware of any datasets that capture a user's detailed preferences in natural language, and attempt to rate recommendations on unseen items.
Existing datasets such as \cite{cpcd,Balog:2019:SIGIR} tend to rely on much simpler characterizations.

\emph{\textbf{Prompting in Large Language Models.}}
Large language models (LLMs) are an expanding area of research with numerous exciting applications. Beyond traditional natural language understanding tasks like summarization, relation mapping, or question answering, LLMs have also proved adept at many tasks such as generating code, generating synthetic data, and multi-lingual tasks \citep{austin2021program, borisov2023language, chowdhery2022palm}. How to prompt these models to generate the best results is a continuing topic of research. Early prompting approaches relied on few-shot prompting, where a small set of training input-output pairs are prepended to the actual input \citep{brown2020language}. Through additional tuning of pre-trained models on tasks described via instructions, LLMs also achieve impressive performance in the zero-shot setting (i.e., models are given a task and inputs, without any previous training examples)~\citep{wei2022finetuned}. %
\citet{p5} test a variety of prompting techniques with a relatively small (less than one billion parameter) LLM trained on a collection of recommendation tasks, finding promising results across multiple tasks and domains, primarily by using item ratings as input.

\section{Experimental Setup}
\label{sec:expsetup}

To study the relationship between item-based and language-based preferences, and their utility for recommendation, we require a parallel corpus from \emph{the same raters} providing both types of information that is \emph{maximally consistent}. There is a lack of existing parallel corpora of this nature, therefore a key contribution of our work is an experiment design that allows such consistent information to be collected. %
Specifically, we designed a two-phase user study where raters were (1) asked to rate items, \emph{and} to describe their preferences in natural language, then (2) recommendations generated based on both types of preferences were uniformly rated by the raters. Hence we perform our experiments in the movie domain, being frequently used for research as movie recommendation is familiar to numerous user study participants.

A key concern in any parallel corpus of this nature is that people may \emph{say} they like items with particular characteristics, but then consume and positively react to quite different items. For instance, this has been observed where people indicate aspirations (e.g.,~subscribe to particular podcasts) yet actually consume quite different items (e.g.,~listen to others)~\citep{Nazari:2022:WWW}.
In general, it has been observed that intentions (such as intending to choose healthy food) often do not lead to actual behaviors \cite{Verplanken:1999:Good}.
Such disparity between corpora could lead to inaccurate prediction about the utility of particular information for recommendation tasks. As such, one of our key considerations was to maximize consistency.

\subsection{Phase 1: Preference Elicitation}
\label{sec:phase1}
\vspace*{-0.25\baselineskip}

Our preference elicitation design collected natural language descriptions of rater interests both at the start and at the end of a questionnaire.
Specifically, raters were first asked to write short paragraphs describing the sorts of movies they liked, as well as the sorts of movies they disliked (free-form text, minimum 150 characters).  These initial liked (+) and disliked (-) self-descriptions for rater $r$ are respectively denoted as $\mathit{desc}^r_{+}$ and $\mathit{desc}^r_{-}$.

Next, raters were asked to name five example items (here, movies) that they like. This was enabled using an online query auto-completion system (similar to a modern search engine) where the rater could start typing the name of a movie and this was completed to specific (fully illustrated) movies. The auto-completion included the top 10,000 movies ranked according to the number of ratings in the MovieLens 25M dataset~\cite{MovieLens} %
to ensure coverage of even uncommon movies. As raters made choices, these were placed into a list which could then be modified.
Each rater was then asked to repeat this process to select five examples of movies they do not like.  These liked (+) and disliked (-) item selections for rater $r$ and item selection index $j \in \{1,
\ldots,5\}$ are respectively denoted as $\mathit{item}^{r,j}_{+}$ and $\mathit{item}^{r,j}_{\mathit{-}}$.

Finally, raters were shown the five liked movies and asked again to write the short paragraph describing the sorts of movies they liked (which we refer to as the \emph{final description}). The was repeated for the five disliked movies. %

\subsection{Phase 2: Recommendation Feedback Collection}
\label{sec:phase2}
\vspace*{-0.25\baselineskip}

To enable a fair comparison of item-based and language-based recommendation algorithms, a second phase of our user study requested raters to assess the quality of recommendations made by a number of recommender algorithms based on the information collected in Phase 1. In particular, past work has observed that completeness of labels is important to ensure fundamentally different algorithms can be compared reliably \cite{Balog:2019:SIGIR, Kaminskas:2016:Diversity}.

\emph{\textbf{Desiderata for recommender selection:}} We aimed for a mix of item-based, language-based, and unbiased recommendations.  
Hence, we collected user feedback (had they seen it or would they see it, and a 1--5 rating in either case) on a shuffled set of 40 movies (displaying both a thumbnail and a short plot synopsis) drawn from four sample pools:
\begin{itemize}
    \item {\bf SP-RandPop}, an unbiased sample of popular items: 10 randomly selected top popular items (ranked 1-1000 in terms of number of MovieLens ratings);
    \item {\bf SP-RandMidPop}, an unbiased sample of less popular items: 10 randomly selected less popular items (ranked 1001-5000 in terms of number of MovieLens ratings);
    \item {\bf SP-EASE}, personalized item-based recommendations: Top-10 from the strong baseline EASE~\cite{ease} collaborative filtering recommender using hyperparameter $\lambda=5000.0$ tuned on a set of held-out pilot data from 15 users; 
    \item {\bf SP-BM25-Fusion}, personalized language-based recommendations: Top-10 from Sparse Review-based Late Fusion Retrieval that, like \cite{Balog:2021:SIGIR}, computes BM25 match between all item reviews in the Amazon Movie Review corpus (v2)~\cite{zemlyanskiy-etal-2021-docent} and rater's natural language preferences ($\mathit{desc}_+$), ranking items by maximal BM25-scoring review.
     
\end{itemize}
Note that SP-RandPop and SP-RandMidPop have 10 different movies for each rater, and that these are a completely unbiased (as they do not leverage any user information, there can be no preference towards rating items that are more obvious recommendations, or other potential sources of bias).  On the other hand, SP-EASE consists of EASE recommendations (based on the user item preferences), which we also evaluate as a recommender---so there is some bias when using this set. We thus refer to the merged set of SP-RandPop and SP-RandMidPop as an {\bf Unbiased Set} in the analysis, with performance on this set being key to our conclusions. 

\subsection{Design Consequences}
Importantly, to ensure a maximally fair comparison of language-based and item-based approaches, consistency of the two types of preferences was key in our data collection approach. As such, we directly crowd-sourced both types of preferences from raters in sequence, with textual descriptions collected twice---before and after self-selected item ratings. This required control means the amount of data per rater must be small. It is also a realistic amount of preference information that may be required of a recommendation recipient in a near-cold-start conversational setting. As a consequence of the manual effort required, the number of raters recruited also took into consideration the required power of the algorithmic comparison, with a key contributions being to the protocol developed rather than data scale. 

Our approach thus contrasts with alternatives of extracting reviews or preference descriptions in bulk from online content similarly to \cite{Bogers:2017:narrative-driven,mysore2023large} (where preferences do not necessarily capture a person's interests fully) and/or relying on item preferences expressed either explicitly or implicitly over time (during which time preferences may change). %
\section{Methods}
\label{sec:methods}
\vspace*{-0.25\baselineskip}

Given our parallel language-based and item-based preferences %
and ratings of 40 items per rater, %
we compare a variety of methods to answer our research questions.  We present the traditional baselines using either item- or language-based preferences, then novel LLM approaches, using items only, language only, or a combination of items and language.

\subsection{Baselines}

To leverage the item and language preferences elicited in Phase 1, we evaluate CF methods as well as a language-based baseline previously found particularly effective \cite{dacrema:2019:progress,Balog:2019:SIGIR}.\footnote{Notably \citet{dacrema:2019:progress} observe that the neural methods do not outperform these baselines.}
Most baseline item-based CF methods use the default configuration in MyMediaLite~\cite{Gantner:2011:MyMediaLite}, including {\bf MostPopular}: ranking items by the number of ratings in the dataset, {\bf Item-kNN}: Item-based k-Nearest Neighbours \cite{Sarwar:2001:ICF}, {\bf WR-MF}: Weighted Regularized Matrix Factorization, a regularized version of singular value decomposition \cite{Hu:2008:CFI}, and {\bf BPR-SLIM}: a Sparse Linear Method (SLIM) that learns a sparse weighting vector over items rated, via a regularized optimization approach \cite{Ning:2011:SSL, Rendle:2009:BBP}. We also compare against our own implementation of the more recent state-of-the-art item-based {\bf EASE} recommender \cite{ease}. %
As a language-based baseline, we compare against {\bf BM25-Fusion}, %
described in Section~\ref{sec:phase2}.
Finally, we also evaluate a random ordering of items in the rater's pool ({\bf Random}) to calibrate against this uninformed baseline.

\subsection{Prompting Methods}

We experiment with a variety of prompting strategies using a variant of the PaLM model (62 billion parameters in size, trained over 1.4 trillion tokens) \cite{chowdhery2022palm}, that we denote moving forward as simply LLM.
Notationally, we assume $t$ is the specific target rater for the recommendation, whereas $r$ denotes a generic rater.  All prompts are presented in two parts: a prefix followed by a suffix which is always the name of the item (movie) to be scored for the target user, denoted as $\langle \mathit{item}^{t}_{*} \rangle$.  The score is computed as the log likelihood of the suffix and is used to rank all candidate item recommendations.\footnote{The full target string scored is the movie name followed by the end-of-string token, which mitigates a potential bias of penalizing longer movie names.} As such, we can evaluate the score given by the LLM to every item in our target set of 40 items collected in Phase 2 of the data collection.

Given this notation, we devise {\bf Completion}, {\bf Zero-shot}, and {\bf Few-shot} prompt templates for the case of {\bf Items only}, {\bf Language only}, and combined {\bf Language+Items} defined as follows:

\subsubsection{Items only}

The completion approach is analogous to that used for the P5 model~\cite{p5}, except that we leverage a pretrained LLM in place of a custom-trained transformer.  The remaining approaches are devised in this work:

\begin{itemize}
    \item {\bf Completion:} $\mathit{item}^{t,1}_{+}$, $\mathit{item}^{t,2}_{+}$, $\mathit{item}^{t,3}_{+}$, $\mathit{item}^{t,4}_{+}$, $\mathit{item}^{t,5}_{+}$,  $\langle \mathit{item}^{t}_{*} \rangle$
    \item {\bf Zero-shot:} I like the following movies: $\mathit{item}^{t,1}_{+}$, $\mathit{item}^{t,2}_{+}$, $\mathit{item}^{t,3}_{+}$, $\mathit{item}^{t,4}_{+}$, $\mathit{item}^{t,5}_{+}$. Then I would also like $\langle \mathit{item}^{t}_{*} \rangle$
    \item {\bf Few-shot ($k$):}
    \begin{tabular}{ll}
    \centeredcell{Repeat $r \in \{1,\ldots,k\}$} \bigg\{ & 
    \centeredcell{User Movie Preferences: $\mathit{item}^{r,1}_{+}$, $\mathit{item}^{r,2}_{+}$, $\mathit{item}^{r,3}_{+}$, $\mathit{item}^{r,4}_{+}$\\
    Additional User Movie Preference: $\mathit{item}^{r,5}_{+}$}
    \end{tabular}\\
    User Movie Preferences: $\mathit{item}^{t,1}_{+}$, $\mathit{item}^{t,2}_{+}$, $\mathit{item}^{t,3}_{+}$, $\mathit{item}^{t,4}_{+}$, $\mathit{item}^{t,5}_{+}$\\
    Additional User Movie Preference: $\langle \mathit{item}^{t}_{*} \rangle$
\end{itemize}

\subsubsection{Language only}
\begin{itemize}
    \item {\bf Completion:} $\mathit{desc}^t_{+}$ $\langle \mathit{item}^{t}_{*} \rangle$
    \item {\bf Zero-shot:} I describe the movies I like as follows: $\mathit{desc}^t_{+}$. Then I would also like $\langle \mathit{item}^{t}_{*} \rangle$
    \item {\bf Few-shot ($k$):}
    \begin{tabular}{ll}
    \centeredcell{Repeat $r \in \{1,\ldots,k\}$} \bigg\{ & 
    \centeredcell{ 
        User Description: $\mathit{desc}^r_{+}$\\
        User Movie Preferences: $\mathit{item}^{r,1}_{+}$, $\mathit{item}^{r,2}_{+}$, $\mathit{item}^{r,3}_{+}$, $\mathit{item}^{r,4}_{+}$, $\mathit{item}^{r,5}_{+}$}
    \end{tabular}\\
    User Description: $\mathit{desc}^t_{+}$\\
    User Movie Preferences: $\langle \mathit{item}^{t}_{*} \rangle$
\end{itemize}

\subsubsection{Language + item}
\begin{itemize}
    \item {\bf Completion:} $\mathit{desc}^t_{+}$ $\mathit{item}^{t,1}_{+}$, $\mathit{item}^{t,2}_{+}$, $\mathit{item}^{t,3}_{+}$, $\mathit{item}^{t,4}_{+}$, $\mathit{item}^{t,5}_{+}$, $\langle \mathit{item}^{t}_{*} \rangle$
    \item {\bf Zero-shot:} I describe the movies I like as follows: $\mathit{desc}^t_{+}$.  I like the following movies: $\mathit{item}^{t,1}_{+}$, $\mathit{item}^{t,2}_{+}$, $\mathit{item}^{t,3}_{+}$, $\mathit{item}^{t,4}_{+}$, $\mathit{item}^{t,5}_{+}$. Then I would also like $\langle \mathit{item}^{t}_{*} \rangle$
    \item {\bf Few-shot ($k$):} 
    \begin{tabular}{ll}
        \centeredcell{Repeat $r \in \{1,\ldots,k\}$} \Bigg\{ &
        \centeredcell{ 
        User Description: $\mathit{desc}^r_{+}$\\
        User Movie Preferences: $\mathit{item}^{r,1}_{+}$, $\mathit{item}^{r,2}_{+}$, $\mathit{item}^{r,3}_{+}$, $\mathit{item}^{r,4}_{+}$\\
        Additional User Movie Preference: $\mathit{item}^{r,5}_{+}$}
    \end{tabular}\\
    User Description: $\mathit{desc}^t_{+}$\\
    User Movie Preferences: $\mathit{item}^{t,1}_{+}$, $\mathit{item}^{t,2}_{+}$, $\mathit{item}^{t,3}_{+}$, $\mathit{item}^{t,4}_{+}$, $\mathit{item}^{t,5}_{+}$\\
    Additional User Movie Preference: $\langle \mathit{item}^{t}_{*} \rangle$
\end{itemize}

\subsubsection{Negative Language Variants}

For the zero-shot cases, we also experimented with negative language variants that inserted the sentences ``I dislike the following movies: $\mathit{item}^{t,1}_{-}$, $\mathit{item}^{t,2}_{-}$, $\mathit{item}^{t,3}_{-}$, $\mathit{item}^{t,4}_{-}$, $\mathit{item}^{t,5}_{-}$'' for {\bf Item} prompts and ``I describe the movies I dislike as follows: $\mathit{desc}^t_{-}$'' for {\bf Language} prompts after their positive counterparts in the prompts labeled {\bf Pos+Neg}.  %

\section{Results}
\label{sec:results}

\subsection{Data Analysis}

\label{sec:times}

We now briefly analyze the data collected from 153 raters as part of the preference elicitation and rating process.\footnote{We recruited 160 raters, but discard those (5) that did not complete both phases of the data collection and those (2) who provided uniform ratings on all item recommendations in Phase 2.}
The raters took a median of 67 seconds to write their initial descriptions summarizing what they like, and 38 seconds for their dislikes (median lengths: 241 and 223 characters, respectively). Providing five liked and disliked items took a median of 174 and 175 seconds, respectively. 
Following this, writing final descriptions of likes and dislikes took a median of 152 and 161 seconds, respectively (median lengths: 205 and 207 characters, respectively). 
We observe that the initial descriptions were produced 3 to 4 times faster than providing 5 example items, in around one minute. 
As we will see below, this difference in effort is particularly pertinent as item-based and description-based recommendation are comparable in performance. A sample of initial descriptions are shown in Table \ref{tab:examples}.

\begin{table}[t]
    \centering
    \caption{Example initial self-descriptions provided by three raters.}
    \vspace*{-0.5\baselineskip}
    \footnotesize
    \label{tab:examples}
    \begin{tabular}{p{0.4cm}|p{7.8cm}|p{6cm}|}%
    \!\!Rater\!\! & \multicolumn{1}{c|}{Liked Movies} & \multicolumn{1}{c|}{Disliked Movies} 
\\
    \hline
    \#1
      & I like comedy movies because i feel happy whenever i watch them. We can watch those movies with a group of people. I like to watch comedy movies because there will be a lot of fun and entertainment. Its very exciting to watch with friends and family.so,I always watch comedy movies whenever I get time. 
      & I am not at all interested in watching horror movies because whenever I feel alone it will always disturb me with the characters in the movie. It will be affected by dreams and mood always. SO, mostly i ignore watching them when i stay alone in the home.
      \\
    \hline
    \#2
      & Fantasy films often have an element of magic, myth, wonder,and the extraordinary. They may appeal to both children and adults, depending upon the particular film. In fantasy films, the hero often undergoes some kind of mystical experience.
      & Horror is scary. I don't like the feeling of being terrified. Some are either sensitive to suspense, gore or frightful images, or they may have had an experience in their life that makes horror seem real.
      \\
    \hline
    \#3
      & I like comedy genre movies, while watching comedy movies I will feel very happy and relaxed. Comedy films are designed to make the audience laugh. It has different kinds of categories in comedy genres such as horror comedy, romantic comedy, comedy thriller,musical-comedy. 
      & I dislike action genre movies because watching fights gives me a headache and bored me. These kinds of movies mainly concentrate on violence and physical feats.
      \\
    \end{tabular}
\end{table}

\begin{table}[t]
    \centering
    \caption{Baseline rating statistics for items in the fully labeled pools of items across all raters.}
    \label{tab:pool_ratings}
    \vspace*{-0.5\baselineskip}
    \begin{tabular}{l|c|c|c:c|}
                 & Movies & Fraction & \multicolumn{2}{c|}{Average Rating}\\
Sample Pool      & Per Rater   &  Seen    & Seen Movies & Unseen Movies \\
\hline
SP-RandPop        & 10 & 22\% & 4.21 & 2.93 \\
SP-RandMidPop     & 10 & 16\% & 4.00 & 2.85 \\
SP-EASE             & 10 & 46\% & 4.51 & 3.16 \\
SP-BM25-Fusion      & 10 & 24\% & 4.38 & 3.11 \\
\hline
SP-Full             & 40 &  27\% & 4.29 & 3.00 \\ %
    \end{tabular}
\end{table}

Next, we analyze the ratings collected for the movies from the four pools described in Section~\ref{sec:expsetup}.  From Table~\ref{tab:pool_ratings}, we observe: (1) The EASE recommender nearly doubles the rate of recommendations that have already been seen by the rater, which reflects the supervised data on which it is trained where raters only rate what they have seen; (2) There is an inherent positive bias to provide a high ratings for movies the rater has already seen as evidenced by the average 4.29 rating in this case; (3) In contrast, the average rating drops to a neutral 3.00 for unseen items.

\subsection{Recommended Items}

\begin{table}[t]
    \centering
    \caption{Main experimental results comparing mean NDCG@10 ($\pm$ 95\% standard error) over raters for all recommendation methods. In each case, the fully judged rater-specific evaluation set is ranked by the given recommendation algorithms. Mean evaluation set sizes are in the first row. Note that performance on the \emph{Unseen} item set is most important in a practical recommendation setting.}
    \label{tab:main_results}
    \vspace*{-0.5\baselineskip}
    \begin{tabular}{l|c|c|c:c|}
               & Full Set & Unbiased Set & \multicolumn{2}{c|}{Items that are} \\
\multicolumn{1}{r|}{Evaluation Set} & SP-Full  & SP-Rand\{Pop,MidPop\} %
& Seen & {\bf Unseen} \\
\hline
\emph{Mean evaluation set size} & \emph{40} & \emph{20} %
& \emph{10.8} & \emph{29.2} \\
\hline
\hline
\multicolumn{5}{l}{Recommendation Algorithm} \\
\hline

Random Baseline	&	0.504 $\pm$ 0.032	&	0.532 $\pm$ 0.034	&	0.876 $\pm$ 0.023	&	0.511 $\pm$ 0.038	\\
Popularity Baseline	&	0.595 $\pm$ 0.032	&	0.624 $\pm$ 0.029	&	0.894 $\pm$ 0.020	&	0.534 $\pm$ 0.036	\\ \hdashline
(Item) EASE	&	{\bf 0.673 $\pm$ 0.038}	&	0.592 $\pm$ 0.030	&	0.899 $\pm$ 0.023	&	0.559 $\pm$ 0.039	\\
(Item) WRMF	&	0.644 $\pm$ 0.036	&	0.644 $\pm$ 0.029	&	0.897 $\pm$ 0.021	&	0.573 $\pm$ 0.037	\\
(Item) BPR-SLIM	&	0.672 $\pm$ 0.037	&	0.617 $\pm$ 0.029	&	{\bf 0.902 $\pm$ 0.021}	&	0.577 $\pm$ 0.037	\\
(Item) KNN Item	&	0.646 $\pm$ 0.038	&	0.610 $\pm$ 0.028	&	0.889 $\pm$ 0.024	&	0.565 $\pm$ 0.037	\\ 
(Language) BM25-Fusion	&	0.519 $\pm$ 0.032	&	0.623 $\pm$ 0.027	&	0.868 $\pm$ 0.023	&	0.542 $\pm$ 0.036	\\ \hdashline
LLM Item Completion	&	0.649 $\pm$ 0.037	&	0.610 $\pm$ 0.027	&	0.889 $\pm$ 0.022	&	0.563 $\pm$ 0.037	\\
LLM Item Zero-shot	&	0.659 $\pm$ 0.037	&	0.631 $\pm$ 0.028	&	0.895 $\pm$ 0.023	&	0.571 $\pm$ 0.037	\\
LLM Item Few-shot (3)	&	0.664 $\pm$ 0.038	&	0.636 $\pm$ 0.027	&	0.897 $\pm$ 0.022	&	0.572 $\pm$ 0.037	\\ \hdashline
LLM Language Completion	&	0.617 $\pm$ 0.032	&	0.617 $\pm$ 0.029	&	0.889 $\pm$ 0.023	&	0.559 $\pm$ 0.035	\\
LLM Language Zero-shot	&	0.612 $\pm$ 0.034	&	0.626 $\pm$ 0.027	&	0.885 $\pm$ 0.024	&	0.563 $\pm$ 0.034	\\
LLM Language Few-shot (3)	&	0.640 $\pm$ 0.036	&	{\bf 0.650 $\pm$ 0.026}	&	0.891 $\pm$ 0.022	&	0.571 $\pm$ 0.038	\\ \hdashline
LLM Item+Language Completion	&	0.654 $\pm$ 0.037	&	0.639 $\pm$ 0.027	&	0.893 $\pm$ 0.022	&	0.568 $\pm$ 0.037	\\
LLM Item+Language Zero-shot	&	0.660 $\pm$ 0.038	&	0.634 $\pm$ 0.028	&	0.897 $\pm$ 0.023	&	{\bf 0.582 $\pm$ 0.037}	\\
LLM Item+Language Few-shot (3)	&	0.663 $\pm$ 0.038	&	0.640 $\pm$ 0.028	&	0.899 $\pm$ 0.022	&	0.570 $\pm$ 0.037	\\ \hdashline
LLM Item Zero-shot Pos+Neg	&	0.647 $\pm$ 0.037	&	0.629 $\pm$ 0.027	&	0.892 $\pm$ 0.023	&	0.569 $\pm$ 0.038	\\
LLM Language Zero-shot Pos+Neg	&	0.612 $\pm$ 0.034	&	0.626 $\pm$ 0.027	&	0.885 $\pm$ 0.024	&	0.563 $\pm$ 0.034	\\
LLM Item+Language Zero-shot Pos+Neg	&	0.662 $\pm$ 0.037	&	0.626 $\pm$ 0.028	&	0.897 $\pm$ 0.023	&	0.577 $\pm$ 0.037	\\
\hline

\hline
    \end{tabular}
\end{table}

Our main experimental results are shown in Table~\ref{tab:main_results}, using NDCG@10 with exponential gain (a gain of 0 for ratings $s<3$ and a gain of $2^{s-3}$ otherwise). We compare the mean performance of various methods using item- and/or language-based preferences (as described in Section~\ref{sec:phase1}) ranking four different pool-based subsets of the 40 fully judged test recommendation items (as described in Section~\ref{sec:phase2}), recalling that the pool for each rater is personalized to that rater. The language-based results use only the initial natural language descriptions, which raters produced much faster than either liked and disliked item choices or final descriptions, yet they yield equal performance to final descriptions.

We begin with general observations.  First, we note the range of NDCG@10 scores within each subset of items is substantially different, due to both the NDCG normalizer that generally increases with a larger evaluation set size, as well as the average rating of each pool.  On the latter note, we previously observed that the subset of {\bf Seen} recommendations in Table~\ref{tab:pool_ratings} has the smallest pool of items and a high positive rating bias that makes it hard to differentiate recommenders on this subset.  However, and as also recently argued in \cite{Pellegrini:2022:Dont}, in a recommendation setting where an item is typically only consumed once (such as movies), we are much more concerned about recommendation performance on the {\bf Unseen} subset vs. the {\bf Seen} subset.  Similarly, we are also concerned with performance on the {\bf Unbiased} set since this subset explores a wide range of popularity and is not biased towards item-based collaborative filtering (CF) methods.

To address our original research questions from Section~\ref{sec:intro}:

{\bf RQ1: Can language-based preferences replace or improve on item-based preferences?}  An initial affirmative answer comes from observing that the LLM Language Few-shot (3) method is competitive with most of the traditional item-based CF methods in this near cold-start setting, which is important since as observed in Section~\ref{sec:times}, language-based preferences took less time to elicit than item-based preferences; furthermore, language-based preferences are transparent and scrutable~\cite{Radlinski:2022:SIGIR}.
    However, there seems to be little benefit to combining language- and item-based preferences as the Item+Language LLM methods do not appear to provide a boost in performance.
    
{\bf RQ2: LLM-based methods vs. CF?} RQ1 has already established that LLM-based methods are generally competitive with item-based CF methods for the Language variants of the LLMs.  However, it should also be noted that in many cases the LLM-based methods can even perform comparatively well to CF methods with only Item-based preferences (i.e., the names of the preferred movies).  A critical and surprising result here is that a pretrained LLM makes a competitive recommender without the large amounts of supervised data used to train CF methods.

{\bf RQ3: Best prompting methodology?} The Few-shot (3) prompting method generally outperforms Zero-shot and Completion prompting methods. 
The difference between Zero-shot and Completion prompting is less pronounced. 
While not shown due to space constraints, increasing the number of Few-shot examples did not improve performance.

{\bf RQ4: Does inclusion of dispreferences help?} In the bottom three rows of Table~\ref{tab:main_results}, we show the impact of including negative item or language preferences for LLM-based recommenders.  There are no meaningful improvements from including both positive and negative preferences (Pos+Neg) over only positive preferences in these LLM configurations.  While not shown due to space constraints, omitting positive preferences and using only negative preferences yields performance at or below the popularity baseline.%

\section{Ethical Considerations}

We briefly consider potential ethical considerations.
First, it is important to consider biases in the items recommended. For instance, it would be valuable to study how to measure whether language-driven recommenders exhibit more or less unintended bias than classic recommenders, such as perhaps preferring certain classes of items over others. Our task was constructed as ranking a fixed corpus of items. As such, all items were considered and scored by the model. Overall performance numbers would have suffered had there been a strong bias, although given the size of our experiments, the existence of bias cannot be ruled out. Larger scale studies would be needed to bound any possible biases present.

Additionally, our conclusions are based on the preferences of a relatively small pool of 153 raters. The small scale and restriction to English-only preferences means we cannot assess whether the same results would be obtained in other languages or cultures.

Finally, we note that the preference data was provided by paid contractors. They received their standard contracted wage, which is above the living wage in their country of employment. 

\section{Conclusion}

In this paper, we collected a dataset containing both item-based and language-based preferences for raters along with their ratings of an independent set of item recommendations.  Leveraging a variety of prompting strategies in large language models (LLMs), this dataset allowed us to fairly and quantitatively compare the efficacy of recommendation from pure item- or language-based preferences as well as their combination.
In our experimental results, we find that zero-shot and few-shot strategies 
in LLMs provide remarkably competitive in recommendation performance for \emph{pure language-based preferences} (no item preferences) in the near cold-start case in comparison to item-based collaborative filtering methods.
In particular, despite being general-purpose, LLMs perform competitively with fully supervised item-based CF methods when leveraging either item-based or language-based preferences. %
Finally, we observe that this LLM-based recommendation approach provides a competitive near cold-start recommender system based on an explainable and scrutable language-based preference representation, thus providing a path forward for effective and novel LLM-based recommenders using language-based preferences.

\bibliographystyle{ACM-Reference-Format}
\bibliography{references}


\begin{thebibliography}{45}


\ifx \showCODEN    \undefined \def \showCODEN     #1{\unskip}     \fi
\ifx \showDOI      \undefined \def \showDOI       #1{#1}\fi
\ifx \showISBNx    \undefined \def \showISBNx     #1{\unskip}     \fi
\ifx \showISBNxiii \undefined \def \showISBNxiii  #1{\unskip}     \fi
\ifx \showISSN     \undefined \def \showISSN      #1{\unskip}     \fi
\ifx \showLCCN     \undefined \def \showLCCN      #1{\unskip}     \fi
\ifx \shownote     \undefined \def \shownote      #1{#1}          \fi
\ifx \showarticletitle \undefined \def \showarticletitle #1{#1}   \fi
\ifx \showURL      \undefined \def \showURL       {\relax}        \fi
\providecommand\bibfield[2]{#2}
\providecommand\bibinfo[2]{#2}
\providecommand\natexlab[1]{#1}
\providecommand\showeprint[2][]{arXiv:#2}

\bibitem[Austin et~al\mbox{.}(2021)]%
        {austin2021program}
\bibfield{author}{\bibinfo{person}{Jacob Austin}, \bibinfo{person}{Augustus
  Odena}, \bibinfo{person}{Maxwell Nye}, \bibinfo{person}{Maarten Bosma},
  \bibinfo{person}{Henryk Michalewski}, \bibinfo{person}{David Dohan},
  \bibinfo{person}{Ellen Jiang}, \bibinfo{person}{Carrie Cai},
  \bibinfo{person}{Michael Terry}, \bibinfo{person}{Quoc Le}, {and}
  \bibinfo{person}{Charles Sutton}.} \bibinfo{year}{2021}\natexlab{}.
\newblock \bibinfo{title}{Program Synthesis with Large Language Models}.
\newblock
\newblock
\showeprint[arxiv]{2108.07732}~[cs.PL]


\bibitem[Balog et~al\mbox{.}(2019)]%
        {Balog:2019:SIGIR}
\bibfield{author}{\bibinfo{person}{Krisztian Balog}, \bibinfo{person}{Filip
  Radlinski}, {and} \bibinfo{person}{Shushan Arakelyan}.}
  \bibinfo{year}{2019}\natexlab{}.
\newblock \showarticletitle{Transparent, Scrutable and Explainable User Models
  for Personalized Recommendation}. In \bibinfo{booktitle}{\emph{Proceedings of
  the 42nd International ACM SIGIR Conference on Research and Development in
  Information Retrieval}} \emph{(\bibinfo{series}{SIGIR '19})}.
  \bibinfo{pages}{265--274}.
\newblock


\bibitem[Balog et~al\mbox{.}(2021)]%
        {Balog:2021:SIGIR}
\bibfield{author}{\bibinfo{person}{Krisztian Balog}, \bibinfo{person}{Filip
  Radlinski}, {and} \bibinfo{person}{Alexandros Karatzoglou}.}
  \bibinfo{year}{2021}\natexlab{}.
\newblock \showarticletitle{On Interpretation and Measurement of Soft
  Attributes for Recommendation}. In \bibinfo{booktitle}{\emph{Proceedings of
  the 44th International ACM SIGIR Conference on Research and Development in
  Information Retrieval}} \emph{(\bibinfo{series}{SIGIR '21})}.
  \bibinfo{pages}{890--899}.
\newblock


\bibitem[Bogers and Koolen(2017)]%
        {Bogers:2017:narrative-driven}
\bibfield{author}{\bibinfo{person}{Toine Bogers} {and} \bibinfo{person}{Marijn
  Koolen}.} \bibinfo{year}{2017}\natexlab{}.
\newblock \showarticletitle{Defining and Supporting Narrative-Driven
  Recommendation}. In \bibinfo{booktitle}{\emph{Proceedings of the Eleventh ACM
  Conference on Recommender Systems}} \emph{(\bibinfo{series}{RecSys '17})}.
  \bibinfo{pages}{238--242}.
\newblock


\bibitem[Borisov et~al\mbox{.}(2023)]%
        {borisov2023language}
\bibfield{author}{\bibinfo{person}{Vadim Borisov}, \bibinfo{person}{Kathrin
  Seßler}, \bibinfo{person}{Tobias Leemann}, \bibinfo{person}{Martin
  Pawelczyk}, {and} \bibinfo{person}{Gjergji Kasneci}.}
  \bibinfo{year}{2023}\natexlab{}.
\newblock \bibinfo{title}{Language Models are Realistic Tabular Data
  Generators}.
\newblock
\newblock
\showeprint[arxiv]{2210.06280}~[cs.LG]


\bibitem[Brown et~al\mbox{.}(2020)]%
        {brown2020language}
\bibfield{author}{\bibinfo{person}{Tom~B. Brown}, \bibinfo{person}{Benjamin
  Mann}, \bibinfo{person}{Nick Ryder}, \bibinfo{person}{Melanie Subbiah},
  \bibinfo{person}{Jared Kaplan}, \bibinfo{person}{Prafulla Dhariwal},
  \bibinfo{person}{Arvind Neelakantan}, \bibinfo{person}{Pranav Shyam},
  \bibinfo{person}{Girish Sastry}, \bibinfo{person}{Amanda Askell},
  \bibinfo{person}{Sandhini Agarwal}, \bibinfo{person}{Ariel Herbert-Voss},
  \bibinfo{person}{Gretchen Krueger}, \bibinfo{person}{Tom Henighan},
  \bibinfo{person}{Rewon Child}, \bibinfo{person}{Aditya Ramesh},
  \bibinfo{person}{Daniel~M. Ziegler}, \bibinfo{person}{Jeffrey Wu},
  \bibinfo{person}{Clemens Winter}, \bibinfo{person}{Christopher Hesse},
  \bibinfo{person}{Mark Chen}, \bibinfo{person}{Eric Sigler},
  \bibinfo{person}{Mateusz Litwin}, \bibinfo{person}{Scott Gray},
  \bibinfo{person}{Benjamin Chess}, \bibinfo{person}{Jack Clark},
  \bibinfo{person}{Christopher Berner}, \bibinfo{person}{Sam McCandlish},
  \bibinfo{person}{Alec Radford}, \bibinfo{person}{Ilya Sutskever}, {and}
  \bibinfo{person}{Dario Amodei}.} \bibinfo{year}{2020}\natexlab{}.
\newblock \bibinfo{title}{Language Models are Few-Shot Learners}.
\newblock
\newblock
\showeprint[arxiv]{2005.14165}~[cs.CL]


\bibitem[Chaganty et~al\mbox{.}(2023)]%
        {cpcd}
\bibfield{author}{\bibinfo{person}{Arun~Tejasvi Chaganty},
  \bibinfo{person}{Megan Leszczynski}, \bibinfo{person}{Shu Zhang},
  \bibinfo{person}{Ravi Ganti}, \bibinfo{person}{Krisztian Balog}, {and}
  \bibinfo{person}{Filip Radlinski}.} \bibinfo{year}{2023}\natexlab{}.
\newblock \showarticletitle{Beyond Single Items: Exploring User Preferences in
  Item Sets with the Conversational Playlist Curation Dataset}. In
  \bibinfo{booktitle}{\emph{Proceedings of the ACM SIGIR Conference on Research
  and Development in Information Retrieval}} \emph{(\bibinfo{series}{SIGIR
  '23})}. \bibinfo{pages}{2754–2764}.
\newblock


\bibitem[Chen et~al\mbox{.}(2022)]%
        {Huiyuan:2022:Denoising}
\bibfield{author}{\bibinfo{person}{Huiyuan Chen}, \bibinfo{person}{Yusan Lin},
  \bibinfo{person}{Menghai Pan}, \bibinfo{person}{Lan Wang},
  \bibinfo{person}{Chin-Chia~Michael Yeh}, \bibinfo{person}{Xiaoting Li},
  \bibinfo{person}{Yan Zheng}, \bibinfo{person}{Fei Wang}, {and}
  \bibinfo{person}{Hao Yang}.} \bibinfo{year}{2022}\natexlab{}.
\newblock \showarticletitle{Denoising Self-Attentive Sequential
  Recommendation}. In \bibinfo{booktitle}{\emph{Proceedings of the 16th ACM
  Conference on Recommender Systems}} \emph{(\bibinfo{series}{RecSys '22})}.
  \bibinfo{pages}{92–101}.
\newblock


\bibitem[Chowdhery et~al\mbox{.}(2022)]%
        {chowdhery2022palm}
\bibfield{author}{\bibinfo{person}{Aakanksha Chowdhery},
  \bibinfo{person}{Sharan Narang}, \bibinfo{person}{Jacob Devlin},
  \bibinfo{person}{Maarten Bosma}, \bibinfo{person}{Gaurav Mishra},
  \bibinfo{person}{Adam Roberts}, \bibinfo{person}{Paul Barham},
  \bibinfo{person}{Hyung~Won Chung}, \bibinfo{person}{Charles Sutton},
  \bibinfo{person}{Sebastian Gehrmann}, \bibinfo{person}{Parker Schuh},
  \bibinfo{person}{Kensen Shi}, \bibinfo{person}{Sasha Tsvyashchenko},
  \bibinfo{person}{Joshua Maynez}, \bibinfo{person}{Abhishek Rao},
  \bibinfo{person}{Parker Barnes}, \bibinfo{person}{Yi Tay},
  \bibinfo{person}{Noam Shazeer}, \bibinfo{person}{Vinodkumar Prabhakaran},
  \bibinfo{person}{Emily Reif}, \bibinfo{person}{Nan Du}, \bibinfo{person}{Ben
  Hutchinson}, \bibinfo{person}{Reiner Pope}, \bibinfo{person}{James Bradbury},
  \bibinfo{person}{Jacob Austin}, \bibinfo{person}{Michael Isard},
  \bibinfo{person}{Guy Gur-Ari}, \bibinfo{person}{Pengcheng Yin},
  \bibinfo{person}{Toju Duke}, \bibinfo{person}{Anselm Levskaya},
  \bibinfo{person}{Sanjay Ghemawat}, \bibinfo{person}{Sunipa Dev},
  \bibinfo{person}{Henryk Michalewski}, \bibinfo{person}{Xavier Garcia},
  \bibinfo{person}{Vedant Misra}, \bibinfo{person}{Kevin Robinson},
  \bibinfo{person}{Liam Fedus}, \bibinfo{person}{Denny Zhou},
  \bibinfo{person}{Daphne Ippolito}, \bibinfo{person}{David Luan},
  \bibinfo{person}{Hyeontaek Lim}, \bibinfo{person}{Barret Zoph},
  \bibinfo{person}{Alexander Spiridonov}, \bibinfo{person}{Ryan Sepassi},
  \bibinfo{person}{David Dohan}, \bibinfo{person}{Shivani Agrawal},
  \bibinfo{person}{Mark Omernick}, \bibinfo{person}{Andrew~M. Dai},
  \bibinfo{person}{Thanumalayan~Sankaranarayana Pillai}, \bibinfo{person}{Marie
  Pellat}, \bibinfo{person}{Aitor Lewkowycz}, \bibinfo{person}{Erica Moreira},
  \bibinfo{person}{Rewon Child}, \bibinfo{person}{Oleksandr Polozov},
  \bibinfo{person}{Katherine Lee}, \bibinfo{person}{Zongwei Zhou},
  \bibinfo{person}{Xuezhi Wang}, \bibinfo{person}{Brennan Saeta},
  \bibinfo{person}{Mark Diaz}, \bibinfo{person}{Orhan Firat},
  \bibinfo{person}{Michele Catasta}, \bibinfo{person}{Jason Wei},
  \bibinfo{person}{Kathy Meier-Hellstern}, \bibinfo{person}{Douglas Eck},
  \bibinfo{person}{Jeff Dean}, \bibinfo{person}{Slav Petrov}, {and}
  \bibinfo{person}{Noah Fiedel}.} \bibinfo{year}{2022}\natexlab{}.
\newblock \bibinfo{title}{PaLM: Scaling Language Modeling with Pathways}.
\newblock
\newblock
\showeprint[arxiv]{2204.02311}~[cs.CL]


\bibitem[Christakopoulou et~al\mbox{.}(2016)]%
        {Christakopoulou:2016:Towards}
\bibfield{author}{\bibinfo{person}{Konstantina Christakopoulou},
  \bibinfo{person}{Filip Radlinski}, {and} \bibinfo{person}{Katja Hofmann}.}
  \bibinfo{year}{2016}\natexlab{}.
\newblock \showarticletitle{Towards Conversational Recommender Systems}. In
  \bibinfo{booktitle}{\emph{Proceedings of the ACM SIGKDD International
  Conference on Knowledge Discovery and Data Mining}}
  \emph{(\bibinfo{series}{KDD '16})}. \bibinfo{pages}{815–824}.
\newblock


\bibitem[Dacrema et~al\mbox{.}(2019)]%
        {dacrema:2019:progress}
\bibfield{author}{\bibinfo{person}{Maurizio~Ferrari Dacrema},
  \bibinfo{person}{Paolo Cremonesi}, {and} \bibinfo{person}{Dietmar Jannach}.}
  \bibinfo{year}{2019}\natexlab{}.
\newblock \showarticletitle{Are We Really Making Much Progress? A Worrying
  Analysis of Recent Neural Recommendation Approaches}. In
  \bibinfo{booktitle}{\emph{Proceedings of the 13th {ACM} Conference on
  Recommender Systems}} \emph{(\bibinfo{series}{RecSys '19})}.
  \bibinfo{pages}{101–109}.
\newblock


\bibitem[Devlin et~al\mbox{.}(2019)]%
        {BERT}
\bibfield{author}{\bibinfo{person}{Jacob Devlin}, \bibinfo{person}{Ming-Wei
  Chang}, \bibinfo{person}{Kenton Lee}, {and} \bibinfo{person}{Kristina
  Toutanova}.} \bibinfo{year}{2019}\natexlab{}.
\newblock \showarticletitle{BERT: Pre-training of Deep Bidirectional
  Transformers for Language Understanding}. In
  \bibinfo{booktitle}{\emph{Proceedings of the 2019 Conference of the North
  {A}merican Chapter of the Association for Computational Linguistics: Human
  Language Technologies, Volume 1 (Long and Short Papers)}}
  \emph{(\bibinfo{series}{NAACL '19})}. \bibinfo{pages}{4171--4186}.
\newblock


\bibitem[Friedman et~al\mbox{.}(2023)]%
        {friedman2023leveraging}
\bibfield{author}{\bibinfo{person}{Luke Friedman}, \bibinfo{person}{Sameer
  Ahuja}, \bibinfo{person}{David Allen}, \bibinfo{person}{Zhenning Tan},
  \bibinfo{person}{Hakim Sidahmed}, \bibinfo{person}{Changbo Long},
  \bibinfo{person}{Jun Xie}, \bibinfo{person}{Gabriel Schubiner},
  \bibinfo{person}{Ajay Patel}, \bibinfo{person}{Harsh Lara},
  \bibinfo{person}{Brian Chu}, \bibinfo{person}{Zexi Chen}, {and}
  \bibinfo{person}{Manoj Tiwari}.} \bibinfo{year}{2023}\natexlab{}.
\newblock \bibinfo{title}{Leveraging Large Language Models in Conversational
  Recommender Systems}.
\newblock
\newblock
\showeprint[arxiv]{2305.07961}~[cs.IR]


\bibitem[Gantner et~al\mbox{.}(2011)]%
        {Gantner:2011:MyMediaLite}
\bibfield{author}{\bibinfo{person}{Zeno Gantner}, \bibinfo{person}{Steffen
  Rendle}, \bibinfo{person}{Christoph Freudenthaler}, {and}
  \bibinfo{person}{Lars Schmidt-Thieme}.} \bibinfo{year}{2011}\natexlab{}.
\newblock \showarticletitle{MyMediaLite: A Free Recommender System Library}. In
  \bibinfo{booktitle}{\emph{Proceedings of the Fifth ACM Conference on
  Recommender Systems}} \emph{(\bibinfo{series}{RecSys '11})}.
  \bibinfo{pages}{305--308}.
\newblock


\bibitem[Gao et~al\mbox{.}(2021)]%
        {gao-convrec-survey}
\bibfield{author}{\bibinfo{person}{Chongming Gao}, \bibinfo{person}{Wenqiang
  Lei}, \bibinfo{person}{Xiangnan He}, \bibinfo{person}{Maarten {de Rijke}},
  {and} \bibinfo{person}{Tat-Seng Chua}.} \bibinfo{year}{2021}\natexlab{}.
\newblock \showarticletitle{Advances and Challenges in Conversational
  Recommender Systems: A Survey}.
\newblock \bibinfo{journal}{\emph{AI Open}}  \bibinfo{volume}{2}
  (\bibinfo{year}{2021}), \bibinfo{pages}{100--126}.
\newblock


\bibitem[Geng et~al\mbox{.}(2022)]%
        {p5}
\bibfield{author}{\bibinfo{person}{Shijie Geng}, \bibinfo{person}{Shuchang
  Liu}, \bibinfo{person}{Zuohui Fu}, \bibinfo{person}{Yingqiang Ge}, {and}
  \bibinfo{person}{Yongfeng Zhang}.} \bibinfo{year}{2022}\natexlab{}.
\newblock \showarticletitle{Recommendation as Language Processing (RLP): A
  Unified Pretrain, Personalized Prompt \& Predict Paradigm (P5)}. In
  \bibinfo{booktitle}{\emph{Proceedings of the 16th ACM Conference on
  Recommender Systems}} \emph{(\bibinfo{series}{RecSys '22})}.
  \bibinfo{pages}{299–315}.
\newblock


\bibitem[Hada and Shevade(2021)]%
        {hada2021rexplug}
\bibfield{author}{\bibinfo{person}{Deepesh~V Hada} {and}
  \bibinfo{person}{Shirish~K Shevade}.} \bibinfo{year}{2021}\natexlab{}.
\newblock \showarticletitle{ReXPlug: Explainable Recommendation using
  Plug-and-Play Language Model}. In \bibinfo{booktitle}{\emph{Proceedings of
  the 44th International ACM SIGIR Conference on Research and Development in
  Information Retrieval}} \emph{(\bibinfo{series}{SIGIR '21})}.
  \bibinfo{pages}{81--91}.
\newblock


\bibitem[Harper and Konstan(2015)]%
        {MovieLens}
\bibfield{author}{\bibinfo{person}{F.~Maxwell Harper} {and}
  \bibinfo{person}{Joseph~A. Konstan}.} \bibinfo{year}{2015}\natexlab{}.
\newblock \showarticletitle{The MovieLens Datasets: History and Context}.
\newblock \bibinfo{journal}{\emph{ACM Transactions on Interactive Intelligent
  Systems}} \bibinfo{volume}{5}, \bibinfo{number}{4}, Article
  \bibinfo{articleno}{19} (\bibinfo{year}{2015}).
\newblock
\showISSN{2160-6455}


\bibitem[He et~al\mbox{.}(2017)]%
        {He:2017:Neural}
\bibfield{author}{\bibinfo{person}{Xiangnan He}, \bibinfo{person}{Lizi Liao},
  \bibinfo{person}{Hanwang Zhang}, \bibinfo{person}{Liqiang Nie},
  \bibinfo{person}{Xia Hu}, {and} \bibinfo{person}{Tat-Seng Chua}.}
  \bibinfo{year}{2017}\natexlab{}.
\newblock \showarticletitle{Neural Collaborative Filtering}. In
  \bibinfo{booktitle}{\emph{Proceedings of the 26th International Conference on
  World Wide Web}} \emph{(\bibinfo{series}{WWW '17})}.
  \bibinfo{pages}{173–182}.
\newblock


\bibitem[Hou et~al\mbox{.}(2022)]%
        {hou2022towards}
\bibfield{author}{\bibinfo{person}{Yupeng Hou}, \bibinfo{person}{Shanlei Mu},
  \bibinfo{person}{Wayne~Xin Zhao}, \bibinfo{person}{Yaliang Li},
  \bibinfo{person}{Bolin Ding}, {and} \bibinfo{person}{Ji-Rong Wen}.}
  \bibinfo{year}{2022}\natexlab{}.
\newblock \showarticletitle{Towards Universal Sequence Representation Learning
  for Recommender Systems}. In \bibinfo{booktitle}{\emph{Proceedings of the
  28th ACM SIGKDD Conference on Knowledge Discovery and Data Mining}}
  \emph{(\bibinfo{series}{KDD '22})}. \bibinfo{pages}{585–593}.
\newblock


\bibitem[Hou et~al\mbox{.}(2023)]%
        {hou2023large}
\bibfield{author}{\bibinfo{person}{Yupeng Hou}, \bibinfo{person}{Junjie Zhang},
  \bibinfo{person}{Zihan Lin}, \bibinfo{person}{Hongyu Lu},
  \bibinfo{person}{Ruobing Xie}, \bibinfo{person}{Julian McAuley}, {and}
  \bibinfo{person}{Wayne~Xin Zhao}.} \bibinfo{year}{2023}\natexlab{}.
\newblock \bibinfo{title}{Large Language Models are Zero-Shot Rankers for
  Recommender Systems}.
\newblock
\newblock
\showeprint[arxiv]{2305.08845}~[cs.IR]


\bibitem[Hu and Yu(2013)]%
        {Hu:2013:Interview}
\bibfield{author}{\bibinfo{person}{Fangwei Hu} {and} \bibinfo{person}{Yong
  Yu}.} \bibinfo{year}{2013}\natexlab{}.
\newblock \showarticletitle{Interview Process Learning for Top-N
  Recommendation}. In \bibinfo{booktitle}{\emph{Proceedings of the ACM
  Conference on Recommender Systems}} \emph{(\bibinfo{series}{RecSys '13})}.
  \bibinfo{pages}{331–334}.
\newblock


\bibitem[Hu et~al\mbox{.}(2008)]%
        {Hu:2008:CFI}
\bibfield{author}{\bibinfo{person}{Yifan Hu}, \bibinfo{person}{Yehuda Koren},
  {and} \bibinfo{person}{Chris Volinsky}.} \bibinfo{year}{2008}\natexlab{}.
\newblock \showarticletitle{Collaborative Filtering for Implicit Feedback
  Datasets}. In \bibinfo{booktitle}{\emph{Proceedings of the 2008 Eighth IEEE
  International Conference on Data Mining}} \emph{(\bibinfo{series}{ICDM
  '08})}. \bibinfo{pages}{263--272}.
\newblock


\bibitem[Jannach et~al\mbox{.}(2021)]%
        {jannach-convrec-survey}
\bibfield{author}{\bibinfo{person}{Dietmar Jannach}, \bibinfo{person}{Ahtsham
  Manzoor}, \bibinfo{person}{Wanling Cai}, {and} \bibinfo{person}{Li Chen}.}
  \bibinfo{year}{2021}\natexlab{}.
\newblock \showarticletitle{A Survey on Conversational Recommender Systems}.
\newblock \bibinfo{journal}{\emph{Comput. Surveys}} \bibinfo{volume}{54},
  \bibinfo{number}{5} (\bibinfo{year}{2021}).
\newblock


\bibitem[Kaminskas and Bridge(2016)]%
        {Kaminskas:2016:Diversity}
\bibfield{author}{\bibinfo{person}{Marius Kaminskas} {and}
  \bibinfo{person}{Derek Bridge}.} \bibinfo{year}{2016}\natexlab{}.
\newblock \showarticletitle{Diversity, Serendipity, Novelty, and Coverage: A
  Survey and Empirical Analysis of Beyond-Accuracy Objectives in Recommender
  Systems}.
\newblock \bibinfo{journal}{\emph{ACM Transactions on Interactive Intelligent
  Systems}} \bibinfo{volume}{7}, \bibinfo{number}{1} (\bibinfo{year}{2016}),
  \bibinfo{pages}{1--42}.
\newblock


\bibitem[Kang et~al\mbox{.}(2017)]%
        {Kang:2017:RecSys}
\bibfield{author}{\bibinfo{person}{Jie Kang}, \bibinfo{person}{Kyle Condiff},
  \bibinfo{person}{Shuo Chang}, \bibinfo{person}{Joseph~A. Konstan},
  \bibinfo{person}{Loren Terveen}, {and} \bibinfo{person}{F.~Maxwell Harper}.}
  \bibinfo{year}{2017}\natexlab{}.
\newblock \showarticletitle{Understanding How People Use Natural Language to
  Ask for Recommendations}. In \bibinfo{booktitle}{\emph{Proceedings of the
  Eleventh ACM Conference on Recommender Systems}}
  \emph{(\bibinfo{series}{RecSys '17})}. \bibinfo{pages}{229--237}.
\newblock


\bibitem[Kang et~al\mbox{.}(2023)]%
        {kang2023llms}
\bibfield{author}{\bibinfo{person}{Wang-Cheng Kang}, \bibinfo{person}{Jianmo
  Ni}, \bibinfo{person}{Nikhil Mehta}, \bibinfo{person}{Maheswaran
  Sathiamoorthy}, \bibinfo{person}{Lichan Hong}, \bibinfo{person}{Ed Chi},
  {and} \bibinfo{person}{Derek~Zhiyuan Cheng}.}
  \bibinfo{year}{2023}\natexlab{}.
\newblock \bibinfo{title}{Do LLMs Understand User Preferences? Evaluating LLMs
  On User Rating Prediction}.
\newblock
\newblock
\showeprint[arxiv]{2305.06474}~[cs.IR]


\bibitem[Liang et~al\mbox{.}(2018)]%
        {Liang:2018:Variational}
\bibfield{author}{\bibinfo{person}{Dawen Liang}, \bibinfo{person}{Rahul~G.
  Krishnan}, \bibinfo{person}{Matthew~D. Hoffman}, {and} \bibinfo{person}{Tony
  Jebara}.} \bibinfo{year}{2018}\natexlab{}.
\newblock \showarticletitle{Variational Autoencoders for Collaborative
  Filtering}. In \bibinfo{booktitle}{\emph{Proceedings of the 2018 World Wide
  Web Conference}} \emph{(\bibinfo{series}{WWW '18})}.
  \bibinfo{pages}{689–698}.
\newblock


\bibitem[Lops et~al\mbox{.}(2011)]%
        {lops2011recsysbook}
\bibfield{author}{\bibinfo{person}{Pasquale Lops}, \bibinfo{person}{Marco
  De~Gemmis}, {and} \bibinfo{person}{Giovanni Semeraro}.}
  \bibinfo{year}{2011}\natexlab{}.
\newblock \showarticletitle{Content-based Recommender Systems: State of the Art
  and Trends}.
\newblock In \bibinfo{booktitle}{\emph{Recommender Systems Handbook}}.
  \bibinfo{publisher}{Springer}, \bibinfo{pages}{73--105}.
\newblock


\bibitem[Malkiel et~al\mbox{.}(2020)]%
        {malkiel2020recobert}
\bibfield{author}{\bibinfo{person}{Itzik Malkiel}, \bibinfo{person}{Oren
  Barkan}, \bibinfo{person}{Avi Caciularu}, \bibinfo{person}{Noam Razin},
  \bibinfo{person}{Ori Katz}, {and} \bibinfo{person}{Noam Koenigstein}.}
  \bibinfo{year}{2020}\natexlab{}.
\newblock \bibinfo{title}{RecoBERT: A Catalog Language Model for Text-Based
  Recommendations}.
\newblock
\newblock
\showeprint[arxiv]{2009.13292}~[cs.IR]


\bibitem[Mysore et~al\mbox{.}(2023a)]%
        {mysore:2023:editable}
\bibfield{author}{\bibinfo{person}{Sheshera Mysore}, \bibinfo{person}{Mahmood
  Jasim}, \bibinfo{person}{Andrew McCallum}, {and} \bibinfo{person}{Hamed
  Zamani}.} \bibinfo{year}{2023}\natexlab{a}.
\newblock \bibinfo{title}{Editable User Profiles for Controllable Text
  Recommendation}.
\newblock
\newblock
\showeprint[arxiv]{2304.04250}~[cs.IR]


\bibitem[Mysore et~al\mbox{.}(2023b)]%
        {mysore2023large}
\bibfield{author}{\bibinfo{person}{Sheshera Mysore}, \bibinfo{person}{Andrew
  McCallum}, {and} \bibinfo{person}{Hamed Zamani}.}
  \bibinfo{year}{2023}\natexlab{b}.
\newblock \bibinfo{title}{Large Language Model Augmented Narrative Driven
  Recommendations}.
\newblock
\newblock
\showeprint[arxiv]{2306.02250}~[cs.IR]


\bibitem[Nazari et~al\mbox{.}(2022)]%
        {Nazari:2022:WWW}
\bibfield{author}{\bibinfo{person}{Zahra Nazari}, \bibinfo{person}{Praveen
  Chandar}, \bibinfo{person}{Ghazal Fazelnia}, \bibinfo{person}{Catherine~M.
  Edwards}, \bibinfo{person}{Benjamin Carterette}, {and}
  \bibinfo{person}{Mounia Lalmas}.} \bibinfo{year}{2022}\natexlab{}.
\newblock \showarticletitle{Choice of Implicit Signal Matters: Accounting for
  User Aspirations in Podcast Recommendations}. In
  \bibinfo{booktitle}{\emph{Proceedings of the ACM Web Conference 2022}}
  \emph{(\bibinfo{series}{WWW '22})}. \bibinfo{pages}{2433–2441}.
\newblock


\bibitem[Ning and Karypis(2011)]%
        {Ning:2011:SSL}
\bibfield{author}{\bibinfo{person}{Xia Ning} {and} \bibinfo{person}{George
  Karypis}.} \bibinfo{year}{2011}\natexlab{}.
\newblock \showarticletitle{SLIM: Sparse Linear Methods for Top-N Recommender
  Systems}. In \bibinfo{booktitle}{\emph{Proceedings of the 2011 IEEE 11th
  International Conference on Data Mining}} \emph{(\bibinfo{series}{ICDM
  '11})}. \bibinfo{pages}{497--506}.
\newblock


\bibitem[Pellegrini et~al\mbox{.}(2022)]%
        {Pellegrini:2022:Dont}
\bibfield{author}{\bibinfo{person}{Roberto Pellegrini}, \bibinfo{person}{Wenjie
  Zhao}, {and} \bibinfo{person}{Iain Murray}.} \bibinfo{year}{2022}\natexlab{}.
\newblock \showarticletitle{Don’t Recommend the Obvious: Estimate Probability
  Ratios}. In \bibinfo{booktitle}{\emph{Proceedings of the 16th ACM Conference
  on Recommender Systems}} \emph{(\bibinfo{series}{RecSys '22})}.
  \bibinfo{pages}{188–197}.
\newblock


\bibitem[Penha and Hauff(2020)]%
        {penha2020does}
\bibfield{author}{\bibinfo{person}{Gustavo Penha} {and}
  \bibinfo{person}{Claudia Hauff}.} \bibinfo{year}{2020}\natexlab{}.
\newblock \showarticletitle{What does BERT know about books, movies and music?
  Probing BERT for Conversational Recommendation}. In
  \bibinfo{booktitle}{\emph{Fourteenth ACM Conference on Recommender Systems}}
  \emph{(\bibinfo{series}{RecSys '20})}. \bibinfo{pages}{388--397}.
\newblock


\bibitem[Radlinski et~al\mbox{.}(2022)]%
        {Radlinski:2022:SIGIR}
\bibfield{author}{\bibinfo{person}{Filip Radlinski}, \bibinfo{person}{Krisztian
  Balog}, \bibinfo{person}{Fernando Diaz}, \bibinfo{person}{Lucas Dixon}, {and}
  \bibinfo{person}{Ben Wedin}.} \bibinfo{year}{2022}\natexlab{}.
\newblock \showarticletitle{On Natural Language User Profiles for Transparent
  and Scrutable Recommendation}. In \bibinfo{booktitle}{\emph{Proceedings of
  the 45th International ACM SIGIR Conference on Research and Development in
  Information Retrieval}} \emph{(\bibinfo{series}{SIGIR '22})}.
  \bibinfo{pages}{2863--2874}.
\newblock


\bibitem[Rendle et~al\mbox{.}(2009)]%
        {Rendle:2009:BBP}
\bibfield{author}{\bibinfo{person}{Steffen Rendle}, \bibinfo{person}{Christoph
  Freudenthaler}, \bibinfo{person}{Zeno Gantner}, {and} \bibinfo{person}{Lars
  Schmidt-Thieme}.} \bibinfo{year}{2009}\natexlab{}.
\newblock \showarticletitle{BPR: Bayesian Personalized Ranking from Implicit
  Feedback}. In \bibinfo{booktitle}{\emph{Proceedings of the Twenty-Fifth
  Conference on Uncertainty in Artificial Intelligence}}
  \emph{(\bibinfo{series}{UAI '09})}. \bibinfo{pages}{452--461}.
\newblock


\bibitem[Rokach and Kisilevich(2012)]%
        {Rokach:2012:Initial}
\bibfield{author}{\bibinfo{person}{Lior Rokach} {and} \bibinfo{person}{Slava
  Kisilevich}.} \bibinfo{year}{2012}\natexlab{}.
\newblock \showarticletitle{Initial Profile Generation in Recommender Systems
  Using Pairwise Comparison}.
\newblock \bibinfo{journal}{\emph{IEEE Transactions on Systems, Man, and
  Cybernetics, Part C (Applications and Reviews)}} \bibinfo{volume}{42},
  \bibinfo{number}{6} (\bibinfo{year}{2012}), \bibinfo{pages}{1854–1859}.
\newblock


\bibitem[Sarwar et~al\mbox{.}(2001)]%
        {Sarwar:2001:ICF}
\bibfield{author}{\bibinfo{person}{Badrul Sarwar}, \bibinfo{person}{George
  Karypis}, \bibinfo{person}{Joseph Konstan}, {and} \bibinfo{person}{John
  Riedl}.} \bibinfo{year}{2001}\natexlab{}.
\newblock \showarticletitle{Item-based Collaborative Filtering Recommendation
  Algorithms}. In \bibinfo{booktitle}{\emph{Proceedings of the 10th
  International Conference on World Wide Web}} \emph{(\bibinfo{series}{WWW
  '01})}. \bibinfo{pages}{285--295}.
\newblock


\bibitem[Sepliarskaia et~al\mbox{.}(2018)]%
        {Sepliarskaia:2018:Optimized}
\bibfield{author}{\bibinfo{person}{Anna Sepliarskaia}, \bibinfo{person}{Julia
  Kiseleva}, \bibinfo{person}{Filip Radlinski}, {and} \bibinfo{person}{Maarten
  de Rijke}.} \bibinfo{year}{2018}\natexlab{}.
\newblock \showarticletitle{Preference Elicitation as an Optimization Problem}.
  In \bibinfo{booktitle}{\emph{Proceedings of the ACM Conference on Recommender
  Systems}} \emph{(\bibinfo{series}{RecSys '18})}. \bibinfo{pages}{172–180}.
\newblock


\bibitem[Steck(2019)]%
        {ease}
\bibfield{author}{\bibinfo{person}{Harald Steck}.}
  \bibinfo{year}{2019}\natexlab{}.
\newblock \showarticletitle{Embarrassingly Shallow Autoencoders for Sparse
  Data}. In \bibinfo{booktitle}{\emph{The World Wide Web Conference}}
  \emph{(\bibinfo{series}{WWW '19})}. \bibinfo{pages}{3251–3257}.
\newblock


\bibitem[Verplanken and Faes(1999)]%
        {Verplanken:1999:Good}
\bibfield{author}{\bibinfo{person}{Bas Verplanken} {and}
  \bibinfo{person}{Suzanne Faes}.} \bibinfo{year}{1999}\natexlab{}.
\newblock \showarticletitle{Good Intentions, Bad Habits, and Effects of Forming
  Implementation Intentions on Healthy Eating}.
\newblock \bibinfo{journal}{\emph{European Journal of Social Psychology}}
  \bibinfo{volume}{29}, \bibinfo{number}{5-6} (\bibinfo{year}{1999}),
  \bibinfo{pages}{591--604}.
\newblock


\bibitem[Wei et~al\mbox{.}(2022)]%
        {wei2022finetuned}
\bibfield{author}{\bibinfo{person}{Jason Wei}, \bibinfo{person}{Maarten Bosma},
  \bibinfo{person}{Vincent~Y. Zhao}, \bibinfo{person}{Kelvin Guu},
  \bibinfo{person}{Adams~Wei Yu}, \bibinfo{person}{Brian Lester},
  \bibinfo{person}{Nan Du}, \bibinfo{person}{Andrew~M. Dai}, {and}
  \bibinfo{person}{Quoc~V. Le}.} \bibinfo{year}{2022}\natexlab{}.
\newblock \bibinfo{title}{Finetuned Language Models Are Zero-Shot Learners}.
\newblock
\newblock
\showeprint[arxiv]{2109.01652}~[cs.CL]


\bibitem[Zemlyanskiy et~al\mbox{.}(2021)]%
        {zemlyanskiy-etal-2021-docent}
\bibfield{author}{\bibinfo{person}{Yury Zemlyanskiy}, \bibinfo{person}{Sudeep
  Gandhe}, \bibinfo{person}{Ruining He}, \bibinfo{person}{Bhargav Kanagal},
  \bibinfo{person}{Anirudh Ravula}, \bibinfo{person}{Juraj Gottweis},
  \bibinfo{person}{Fei Sha}, {and} \bibinfo{person}{Ilya Eckstein}.}
  \bibinfo{year}{2021}\natexlab{}.
\newblock \showarticletitle{{DOCENT}: Learning Self-Supervised Entity
  Representations from Large Document Collections}. In
  \bibinfo{booktitle}{\emph{Proceedings of the 16th Conference of the European
  Chapter of the Association for Computational Linguistics: Main Volume}}
  \emph{(\bibinfo{series}{EACL '21})}. \bibinfo{pages}{2540--2549}.
\newblock


\end{thebibliography}

\end{document}